\documentclass[]{aastex701}
\usepackage{amsmath}
\newcommand{\eb}{\begin{equation}}
\newcommand{\ee}{\end{equation}}

\newcommand{\uasyr}{\mbox{$\mu$as yr$^{-1}$}}
\newcommand{\uas}{\mbox{$\mu$as}}

\definecolor{rkka}{RGB}{219,66,32}
\definecolor{151}{rgb}{0.1,0.5,0.1}
\shorttitle{A posteriori correction of Gaia CRF proper motions}
\shortauthors{Valeri V Makarov}

\begin{document}

\title{A posteriori correction of Gaia CRF proper motions using conditional probability\\ and zero-parallax prior
}

\author[0000-0003-2336-7887]{Valeri V.\ Makarov}
\email{valeri.makarov@gmail.com}
\affiliation{U.S. Naval Observatory, 3450 Massachusetts Ave NW, Washington, DC 20392-5420, USA\\
}
\correspondingauthor{Valeri V. Makarov}
\begin{abstract}
The formal precision of Gaia CRF proper motions, and in principle positions, can be statistically improved by introducing conditional probability based on the zero-parallax prior for these extremely distant sources. The method developed in this study produces statistically conditioned proper-motion estimates under the assumed per-source Gaussian covariance model. It is demonstrated that thus corrected proper motion vectors of 1.2 million CRF objects in Gaia DR3 show a tighter distribution around zero and a general reduction in the rate of outliers by approximately 15\%. A general vector spherical harmonic fit to degree 4 is produced with the corrected data for a filtered subset of CRF quasars with redshifts between 1 and 1.5. This application, predicated on the unknown stochastic component of the global parallax field, shows a 3-sigma change in the estimated magnitude of the Galactocentric acceleration of the solar system barycenter, quantifying the sensitivity of this fundamental determination to the proposed conditioning of proper motions.
\medskip
\begin{flushright}
    \itshape In memory of Prof. Tateosh Agekian
\end{flushright}

\end{abstract}


\section{Introduction} \label{int.sec}
The collection of Gaia Celestial Reference Frame (CRF) objects\footnote{The list of cross-matched sources is available at \url{https://vizier.u-strasbg.fr/viz-bin/VizieR-3?-source=I/355/gcrf3xm}} is an important subset of the Gaia space mission \citep{2016A&A...595A...1G} main catalog. As explained by \citet{2022A&A...667A.148G}, the main purpose of Gaia CRF3 is to remove the intrinsic 6-rank degeneracy of the global astrometric solution, which is based on differential measurements of arcs between the observed sources. This task is dual, since the six degrees of freedom include three rotations of the coordinate system and three constant spins. While the former triplet of transformations is a matter of practical convenience, the undetermined spin, as reflected in the proper motion system, has important physical meaning. Here, a strong ``prior" should be used to tie up the measured angular motion of billions of Gaia sources within a certain acceptable concept. This concept is essentially the Mach's principle, which states that a local inertial frame is defined by the entire ensemble of distant physical bodies \citep{RevModPhys.36.510}. One can remove the 3-rank deficiency of the spin by setting the average rotation of the proper motion system to zero. This requires a large sample of globally distributed, very distant sources. The spin vector of the CRF sample can be computed and simply subtracted from the estimated proper motions for the entire collection of Gaia objects. Obviously, this crucial step has a bearing on such important applications as Galactic rotation, determination of Oort parameters, the physical motion of the Local Group.

The five astrometric parameters determined by Gaia for a given source can be viewed as a random vector drawn from a multivariate distribution, which has a known (or estimated) dispersion matrix (also called variance-covariance matrix) but unknown mean values. Inference about the means, which are equal to the true parameters for symmetric and unbiased distributions of measurement error, is obtained from the sample vector, which is limited to one realization. 
Trivially, the sample mean estimate is the measured value, and a catalog user can do nothing to improve the precision of this estimation. The situation is different when some to the multivariate components are known a priori. For the sources in the Gaia CRF collection, which are extremely distant quasars and AGNs, the true parallax is practically zero. Calculations of the expected annual parallax of a source in the standard $\Lambda$CDM cosmology with the cosmological parameters determined by the Planck mission \citep{2020A&A...641A...6P} provide 0.023 \uas\ at redshift 0.01, 0.0023 \uas\ at redshift 0.1, 0.0005 \uas\ at redshift 0.5, for example, which are vanishingly small numbers compared to the parallax measurement error. Therefore, the given parallax measurement does not represent the true value at all, but in fact is just the sample realization of the measurement error. This value can be used to improve the estimates of the other parameters, whose true means should still be inferred.

If the astrometric parameter measurements are independent, there is no statistical relation between them, and each parameter can be treated as a univariate variable. In reality, since the estimates are obtained in a global least-squares adjustment, the measurements are correlated \citep{2012A&A...543A..14H}. In particular, a given parallax measurement represents a single random sample of the population with a zero mean. If, for example, the measured parallax of a quasar is 200 \uas\ and the correlation coefficient of parallax and another parameter $x$ is $+0.5$, the measured $x$ {\it is likely} to have a positive measurement error with respect to its true mean. To develop this useful observation in quantitative terms, we have to employ the classical conditional probability paradigm. Assuming that the measurement errors follow a multivariate normal distribution (a basic assumption that has already been used to derive the astrometric solution), we obtain the following corrected estimate (Appendix A):
\eb 
\boldsymbol{\hat\mu}\equiv \boldsymbol{\mu}-E(\mu|\varpi)=\boldsymbol{\mu}-\frac{\varpi}{\sigma_\varpi}[r_{01}\,\sigma_1,r_{02}\,\sigma_2]^T,
\label{mu.eq}
\ee 
where $\boldsymbol{\mu}$ is the measured proper motion value for a given object, $\varpi$ is the measured parallax, $\sigma_\varpi$ is the parallax formal error, $\sigma_1$ and $\sigma_2$ are the formal errors of the proper motion components ($\mu_{\alpha*}$ and $\mu_\delta$, respectively), and $r_{01}$ and $r_{02}$ are the corresponding correlation coefficients. Note that the correction reverses the sign of the correlation coefficients, which is intuitively clear from general geometric considerations. 

A similar equation holds for the position components $\boldsymbol{p}$ if improved positions are required for a specific application, for example, computing the radio-optical position offsets for radio-loud quasars in ICRF3 \citep[e.g.,][]{2023AJ....166....8M}. Vector $\boldsymbol{\mu}$ is then replaced with a vector of position offset $\boldsymbol{p}$ in the local tangential frame, and the correlation coefficients $r_{01}$ and $r_{02}$ are the corresponding parallax-position correlations from the given $5\times 5$ formal covariance matrix. Finally, some tasks may require all four parameters corrected in a self-consistent way. For stars, epoch transformation of positions is the most common task of this kind, although this transformation is not recommended for CRF quasars. Eq. \ref{mu.eq} is easily expandable for a 4D correction by replacing $\boldsymbol{\mu}$ with a 4D vector $\nu$ comprising the position and proper motion components, and engaging all four formal correlations.

Let us now consider the statistical meaning of Eq. \ref{mu.eq}. It tells us that the likeliest value of true proper motion $\hat{\boldsymbol{\mu}}$ given the measured parallax value $\varpi$ is not equal to the measured $\boldsymbol{\mu}$ but to a shifted value by the conditional mean, which is the expectation of the conditional sample. The zero-parallax condition changes the expectation whenever the correlation between these components is nonzero. The conditional dispersion is also different from the full-rank covariance matrix given in the catalog (Appendix A), which provides the following formula for the parallax-corrected covariance matrix of the proper motion vector:
\eb 
\hat{\boldsymbol{\Sigma}}_\mu=\boldsymbol{\Sigma}_\mu -
\begin{bmatrix} r_{01}^2\sigma_1^2 & r_{01}r_{02}\sigma_1\sigma_2 \\ r_{01}r_{02}\sigma_1\sigma_2 & r_{02}^2\sigma_2^2 \end{bmatrix}.
\label{sigma.eq}
\ee 
Note that the correction term in square brackets, $\boldsymbol{\Sigma}_{\mu\varpi}\boldsymbol{\Sigma}_{\varpi\varpi}^{-1}\boldsymbol{\Sigma}_{\varpi\mu}$, is positive semidefinite---rank one in the two-dimensional proper-motion case, and the zero matrix when both parallax--proper-motion correlations vanish (which does not occur in practice for the present sample). Accordingly, the conditionally corrected dispersion is not larger than the catalogued covariance in the positive-semidefinite (Loewner) sense, and it is unchanged when the relevant correlations vanish.

The per-source covariance matrix published in DR3 is
derived from the local AGIS normal-equation block for each source, marginalized
over the global calibration and attitude parameters in the standard block-wise
elimination procedure \citep[see][]{2012A&A...543A..14H}.  To the extent that this
marginalisation is exact, the published $5\times5$ covariance matrix does encode the propagated
uncertainty from all higher-level degrees of freedom, including attitude and
calibration.  Correlations between the errors of different sources that share common attitude
or calibration residuals are not represented in the available covariances.  Inter-
source covariances are not required for computing the individual conditional corrections per Eq.~\ref{mu.eq}, but they may affect
the interpretation of global quantities derived from the corrected catalogue, including VSH coefficients of radio-optical CRF offsets, spin, glide, and secular-aberration terms.

\section{Does it work?}
To validate the conditional zero-parallax correction of proper motions, I used the Gaia data for 1.2 million CRF3 objects with known spectroscopic or Machine Learning-predicted redshifts from the sample constructed in \citep{2025NatAs...9.1396M}. Only sources with nominal 5-parameter solutions have been used in this analysis. The $\sim0.4$ million 6-parameter solutions (flag 91 in {\tt astrometric\_params\_solved} field of Gaia catalog) were filtered out because of their inferior astrometric quality and higher covariance dimensionality. 

\begin{figure*}
    \includegraphics[width=0.45 \textwidth]{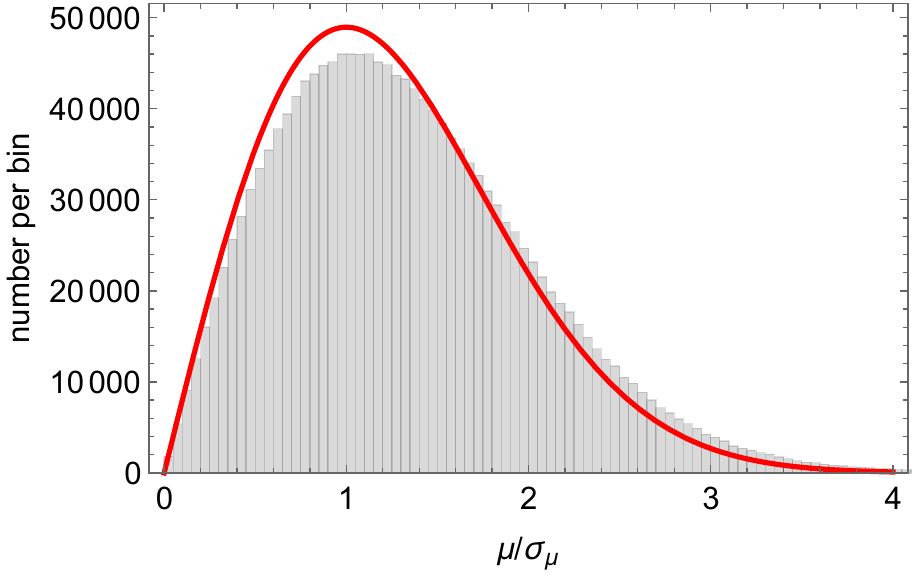}
    \includegraphics[width=0.45 \textwidth]{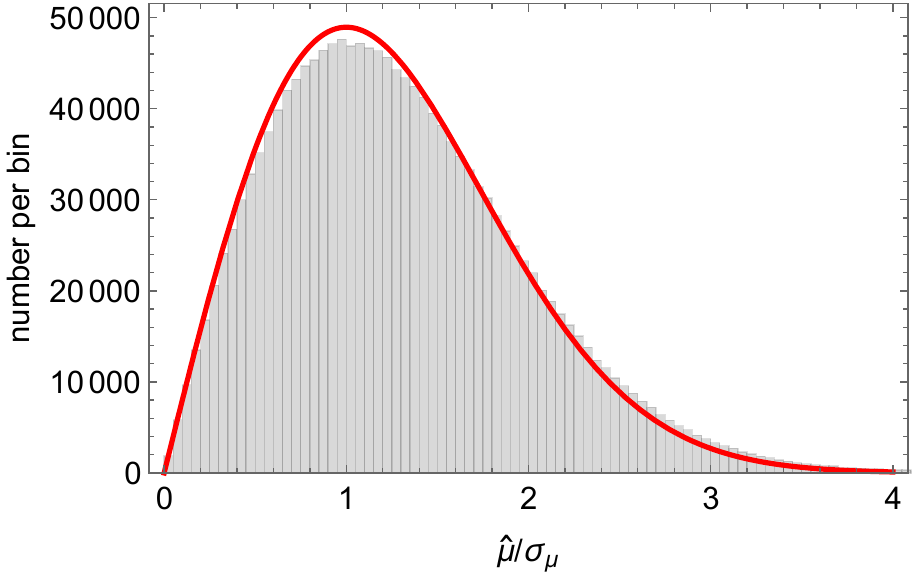}
    \caption{Sample distributions of original (left plot) and conditionally corrected (right  plot) Gaia CRF error-normalized proper motion magnitudes. The solid line in both cases represents the expected probability density of $\chi(2)$ distribution rescaled to the given size of the sample. The normalization of the corrected proper motions in the right plot is performed with the original formal covariances.} 
    \label{mu.fig}
\end{figure*}

Fig. \ref{mu.fig}, left panel, shows the histogram of 1.2 million Gaia CRF3 proper motion magnitudes normalized by their formal errors \citep{2022ApJ...933...28M}. This is essentially the square root of the computed $\chi^2(2)$ value, which is a 2D analog of the commonly used $z$-score. The red solid line represents the theoretically expected probability density curve of the $\chi$-distribution with 2 degrees of freedom, scaled to the given sample size.\footnote{The sample distribution of CRF proper motion magnitudes should follow the $\chi(2)$ curve within the assumed statistical model because the true proper motions of quasars at cosmological distances are vanishingly small} We can see that the empirical histogram deviates from the theoretical curve, in that there is a deficit of data points around the peak and an excess of points in the tail of the distribution. The flatter empirical distribution can be interpreted as a moderate underestimation of the formal proper motion uncertainties or, more radically, as a departure from the assumed Gaussianity of measurement errors. In reality, the measured proper motion components are also perturbed by the random realizations of parallax, and we can further improve them by subtracting the conditional mean.

Using Eq. \ref{mu.eq}, corrected proper motions were computed directly from the data given in the main Gaia catalog. Most of the updates prove to be quite small because the correlations $r_{01}$ and $r_{02}$ tend to be close to zero in absolute value, and the formal uncertainty of parallax $\sigma_\varpi$ is usually larger than the uncertainty of proper motion components. The histograms of the correction components are sharply peaked at zero but have symmetric shallow tails extending to high values. The corrections are systematically larger in RA than in Dec, with the median absolute values 46 and 42 \uasyr, respectively. The 0.9 quantiles of the absolute corrections are 310 and 278 \uasyr, and the 0.99 quantiles are 1130 and 1014 \uasyr. Relative to the formal errors of proper motions given in the catalog, the absolute correction values are greater than 0.1 for 48\% of the sample, and greater than 0.5 for 7\% of the sample. No significant correlations were found between the conditional proper motion corrections and $G$ magnitude, BP$-$RP color, {\tt ra\_rec\_corr}, or {\tt phot\_bp\_rp\_excess\_factor} parameters.

Despite the commonly small absolute and relative values of conditional corrections, we can see a remarkable improvement in the resulting distribution of standardized magnitudes in Fig. \ref{mu.fig}, right panel. We used the original formal covariances for this normalization, so that the original and corrected histograms can be directly compared. The corrected distribution is much closer to the expected $\chi(2)$ curve, and the rate of outliers in the tail is reduced. Quantitatively, the improvement in the rate of outliers is estimated by counting the data points outside of the $\mu/\sigma_\mu=3$ threshold at 15\% (dropping from 0.0223 to 0.0190).

To clarify the issue of possibly underestimated proper motion errors in Gaia or non-Gaussianity of their distribution, we have to consistently compare the corrected vectors with corrected uncertainties by Eq. \ref{sigma.eq}. Remarkably, the conditionally corrected normalized magnitudes $\mu/\sigma_\mu$ also show significant improvement in the rate of outliers. Specifically, the fraction of outliers above 3 reduced from 0.0223 to 0.0193, and that above 4 from 0.0025 to 0.0016, i.e., by 13\% and 36\%, respectively.
With these significant improvements, the rates are still higher than the theoretically expected 0.0111 and 0.0003. The persistent tails of the distribution may reflect a mixture of stellar interlopers, source-structure or photocentre effects in AGNs, blending or crowding, and residual Gaia systematics not captured by the per-source covariance model \citep{2019ApJ...885L...4S, 2022NatAs...6.1185M, 2024A&A...692A.154W,2026arXiv260610655W}.

\begin{figure*}
    \includegraphics[width=0.32 \textwidth]{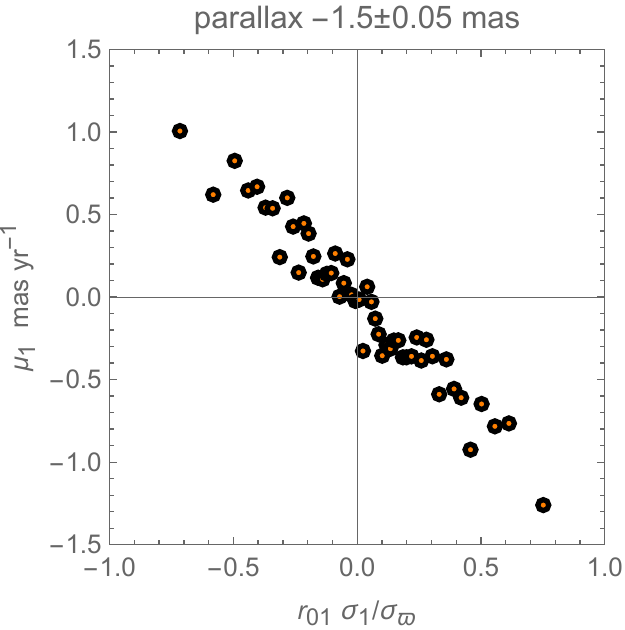}
    \includegraphics[width=0.32 \textwidth]{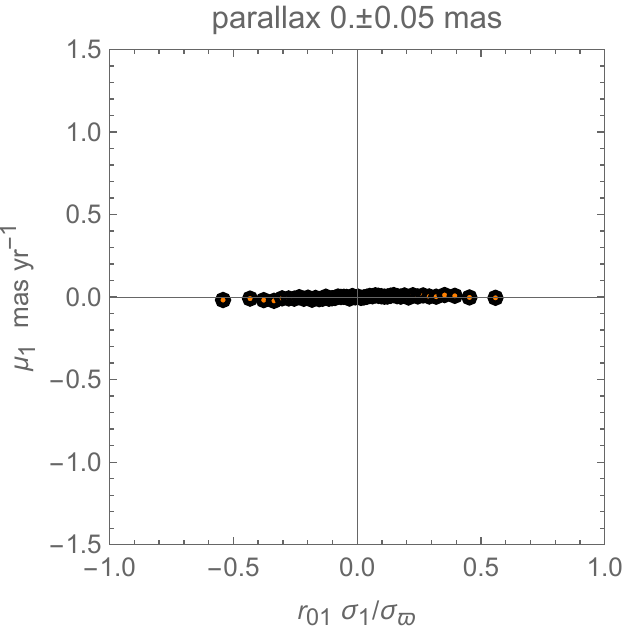}
    \includegraphics[width=0.32 \textwidth]{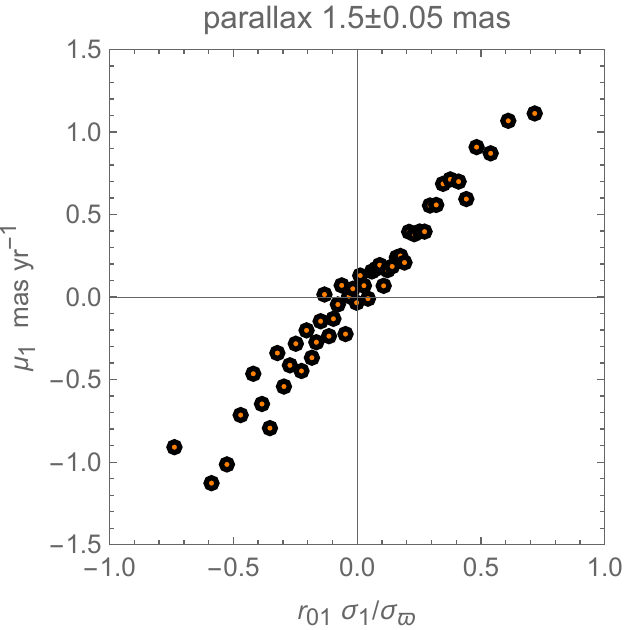}
    \caption{Empirical dependencies of median proper motion in RA on the construct of formal covariance parameters $r_{01}\sigma_1/\sigma_{\varpi}$ entering Eq.~\ref{mu.eq} for three subsets of Gaia CRF3 sources with parallaxes close to $-1.5$ mas (left), $0$ mas (center), and $+1.5$ mas (right). The slope of the linear trends is consistent with the conditioning of proper motions per Eq.~\ref{mu.eq}.}
    \label{cond.fig}
\end{figure*}

As an internal sanity check, the correction was recomputed after randomly permuting the parallaxes and their formal errors among the sources while keeping all other data fixed. This destroys the physical parallax--proper-motion association of each source and, as expected, worsens the error-normalized distribution: the outlier fraction above 3 rises from 0.0223 to 0.0266 (Table~\ref{outlier.tab}). The test confirms that the improvement seen with the true parallaxes is not an artifact of the algebraic procedure and provides a check on the signs adopted in Eqs.~\ref{mu.eq} and \ref{sigma.eq}. The outlier fractions for all cases are collected in Table~\ref{outlier.tab}. Finally, Fig. \ref{cond.fig} provides direct internal evidence that the conditional, parallax-driven errors are real and present in Gaia CRF3 data. The plots show the binned median RA proper motion component ($\mu_1$ in Eq. \ref{mu.eq}) as function of constructs $r_{01}\sigma_1/\sigma_{\varpi}$ for three subsets of the CRF sample: measured parallaxes are within $-1.5\pm0.05$ (8229 objects), $0\pm0.05$ (151507 objects), and $+1.5\pm0.05$ (7633 objects). Each subset is divided into 50 contiguous bins by the sorted construct. Note that only the per-source data as given in the Gaia catalog are used for these plots without any modifications. The slopes of the linear trends expected from Eq. \ref{mu.eq} are consistent with the fixed parallax values.

\begin{table}[ht]
\centering
\caption{Fraction of error-normalized proper-motion magnitudes exceeding the $3\sigma$ and $4\sigma$ thresholds for the 1.2 million 5-parameter CRF sources.}
\label{outlier.tab}
\begin{tabular}{lcc}
\hline\hline
Case & $f(>3\sigma)$ & $f(>4\sigma)$ \\
\hline
Original $\boldsymbol{\mu}$, original $\boldsymbol{\Sigma}_\mu$ & 0.0223 & 0.0025 \\
Corrected $\hat{\boldsymbol{\mu}}$, original $\boldsymbol{\Sigma}_\mu$ (Eq.~\ref{mu.eq}) & 0.0190 & 0.0022 \\
Corrected $\hat{\boldsymbol{\mu}}$, corrected $\hat{\boldsymbol{\Sigma}}_\mu$ (Eqs.~\ref{mu.eq},~\ref{sigma.eq}) & 0.0193 & 0.0016 \\
Permuted-parallax null test & 0.0266 & 0.0030 \\
\hline
Theoretical $\chi(2)$ & 0.0111 & 0.0003 \\
\hline
\end{tabular}
\end{table}

\section{Determination of local acceleration of free fall with corrected CRF proper motions}
\label{3.sect}
The solar system barycenter is in free fall as it accelerates in the direction of local gravitational potential gradient in the Galaxy. This direction is expected to be well aligned with the Galactic center, apart from a possible vertical component towards the concentration of mass in the disk. The vertical acceleration is expected to be relatively small due to the current position of the Sun close to the plane at roughly 1/10 of the oscillation amplitude. The main galactocentric component of acceleration is too small ($\sim 2 \times10^{-10}$ m s$^{-2}$) to be measured directly via reflex angular acceleration of distant stars, for example, but it causes an observable secular aberration drift \citep{2003A&A...404..743K, 2006AJ....131.1471K, 2011A&A...529A..91T}. The astrometric effect can be viewed as a slowly rotating dipole of shifted positions, which generates a dipole perturbation of angular proper motions of all celestial sources aligned with the instantaneous acceleration vector. Since aberration of light is an absolute special relativity effect, the current acceleration includes contributions from external gravitational potentials (such as a possible acceleration of the Milky Way toward Andromeda, acceleration of the Local Group toward the Virgo Supercluster, etc.), but these components are negligible. 

Estimation of the solar system acceleration \citep{2021A&A...649A...9G} is one of the main achievements of the Gaia mission. Since the sought signal is approximately 1/100 of the median CRF proper motion uncertainty, a sample of at least $10^5$ sources is required to boost the signal-to-noise ratio. Fortunately, the expected dipole signal is nicely separated from the discussed intrinsic degeneracy of Gaia astrometry with respect to rotation and spin. Indeed, if we represent the observed field of CRF proper motions as a series of vector spherical harmonic (VSH) functions,
\eb 
\boldsymbol{\mu}(l,b)=\sum_{\{k,p,q\}}\left(a_{\{k,p,q\}}^{(E)}\boldsymbol{E}_{\{k,p,q\}}(l,b)+a_{\{k,p,q\}}^{(M)}\boldsymbol{M}_{\{k,p,q\}}(l,b)\right),
\label{vsh.eq}
\ee 
where ${\{k,p,q\}}$ are certain combination of indices called kind, degree, and order, respectively, we find that the indefinite spin is fully expressed with the three first-degree ($p=1$) magnetic VSH $\boldsymbol{M}$ terms, while the aberration signal is contained within the triplet of electric VSH $\boldsymbol{E}$ of first degree. An exact representation generally requires an infinite series, so that $p=1,2,\ldots,\infty$. For a discretized vector field (a set of tangent vectors with origins at specific spherical coordinates $(l,b)$), the series should be truncated, and Eq. \ref{vsh.eq} becomes approximate. It is customary to include all VSH terms up to a certain degree $P$. The VSH decomposition can be computed in any coordinate system, but the galactic system $(l,b)$ is a natural choice since the signal is concentrated in a single \{ele,1,1,1\}$\equiv \boldsymbol{E}_{\{1,1,1\}}$ term.\footnote{For the definition and nomenclature of VSH used in this paper, see \citep{2025NatAs...9.1396M}} The amplitude of the  \{ele,1,1,1\} dipole as determined by \citet{2021A&A...649A...9G} is $5.1\pm 0.4$ \uasyr. Follow-up analyses \citep[e.g., ][]{2025arXiv250802810T} produced broadly consistent estimates, but also revealed a borderline significant variation of this parameter with redshift (which is inconsistent with the secular aberration model) and the presence of totally unexpected redshift-dependent magnetic rotation terms of first degree. 

One may inquire if some of these problems are caused by the statistical relation between the measured proper motions and parallaxes. Gaia parallaxes of distant CRF objects appear to have a complex, sky-correlated pattern of systematic error or bias \citep{2021A&A...649A...4L}. It is yet to be understood if this mostly negative bias (also called zero-point in the literature) originates from deterministic instrumental effects or has purely stochastic nature. In the space of spherical harmonics, errors of global astrometric solutions can be both systematic and accidental \citep{2012AJ....144...22M}. The smooth component of the parallax bias distribution on the sphere can include a yet unknown contribution from the purely random measurement error, which propagates unevenly in the degrees of spatial frequency of a given single realization (global astrometric solution). The power spectrum of error propagation is red at low VSH degrees, the accidental part of the parallax offset is significant, and an interesting possibility emerges to improve the estimates of the VSH proper motion fit, including the magnitude of the \{ele,1,1,1\} dipole, by subtracting the correlated part of the error from each individual proper motion vector.

\begin{figure}
    \includegraphics[width=0.95\textwidth]{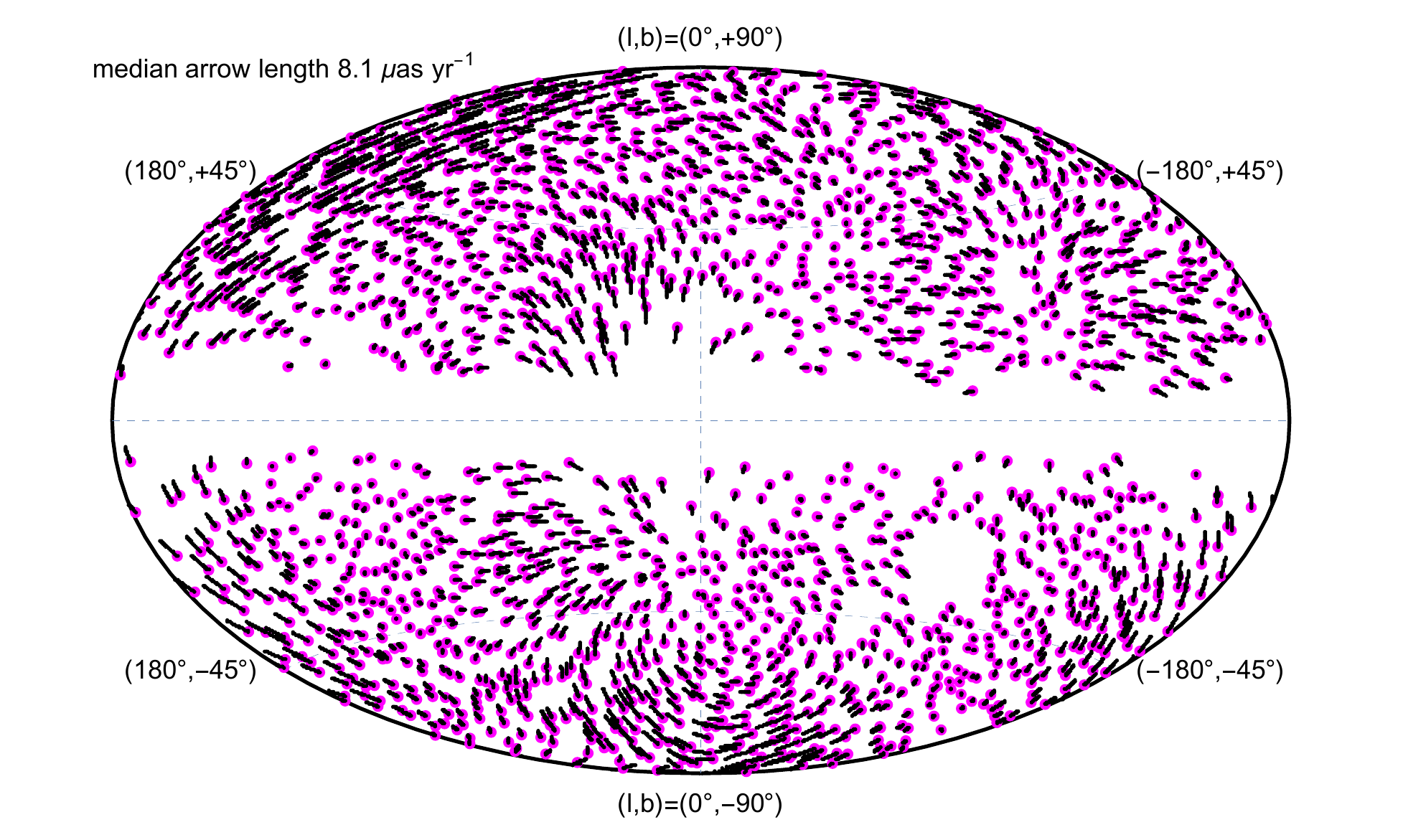}
   \caption{VSH-fitted proper motion field of Gaia CRF quasars with ML-predicted redshifts $1.0<z<1.51$ on the celestial sphere. Graphical presentation in the Aitoff Galactic projection with the Galactic center direction at the center of the plot. Magenta dots at the origin of vectors indicate the mean positions of sources. Only 0.7\% of sources in this sample are shown. The length and direction of small vectors represent the cumulative fit for the corresponding quasar from the 48 VSH terms to degree 4. The median length of the arrows is 8.1 \uasyr.}
    \label{field1.fig}
\end{figure}

A pilot investigation in this direction was made using the sample of Gaia CRF objects with spectroscopic or machine-learning synthetic redshifts from \citep{2025NatAs...9.1396M}. The test sample included 284,639 sources with redshifts between 1 and 1.5 for verification with previous results. The fit included 48 VSH terms to degree $P=4$. The first fit used the original proper motions (converted to the galactic coordinates) and their formal covariances for weights in the weighted global adjustment. The resulting 48 coefficients $a_{\{k,p,q\}}^{(M)}$ and $a_{\{k,p,q\}}^{(E)}$ are mostly within 1--2 times their formal uncertainties, with the exception of \{mag,2,1,1\} rotation, \{ele,1,1,1\} dipole, \{ele,1,2,2\} quadrupole, \{mag,1,3,1\}, and \{ele,2,4,2\}, which are above 3-sigma in absolute value. The dipole term is the one we are mostly interested in. Its fitted coefficient corresponds to an amplitude of $3.35\pm 0.72$ \uasyr (Table \ref{zp.tab}), which is below the Gaia estimate by $2.4\sigma$. The fitted field in graphical form is shown in Fig. \ref{field1.fig}. We note that the expected galactocentric dipole is not conspicuous because it is mixed with the other large terms.

All individual proper motion vectors and their covariances were then corrected using Eqs. \ref{mu.eq} and \ref{sigma.eq}. With this updated information, another weighted VSH adjustment was performed for the same sample of quasars. The design matrix could be reused, but the weight matrix and the right-hand part of the condition equations had to be recomputed. For most of the 48 coefficients, the numerical changes are within their formal errors, and the initially insignificant terms remained insignificant except for one term, \{ele,0,1,0\}, which is a zonal polar dipole. The coefficient for this term increased from $-1.0$ to $-7.8$ with a formal error of 1.8 \uasyr. Surprisingly, the largest terms \{mag,2,1,1\} and \{ele,1,1,1\} both increased in absolute value. From the latter, the corrected amplitude of the galactocentric dipole is $5.59\pm 0.73$ \uasyr, which is closer to the Gaia-determined value representing a $3\sigma$ change. This agreement should not be interpreted as an independent validation of the corrected dipole, but rather as evidence that low-degree VSH estimates are sensitive to the parallax--proper-motion covariance structure and to the sky distribution of parallax offsets.

\begin{figure}
    \includegraphics[width=0.95\textwidth]{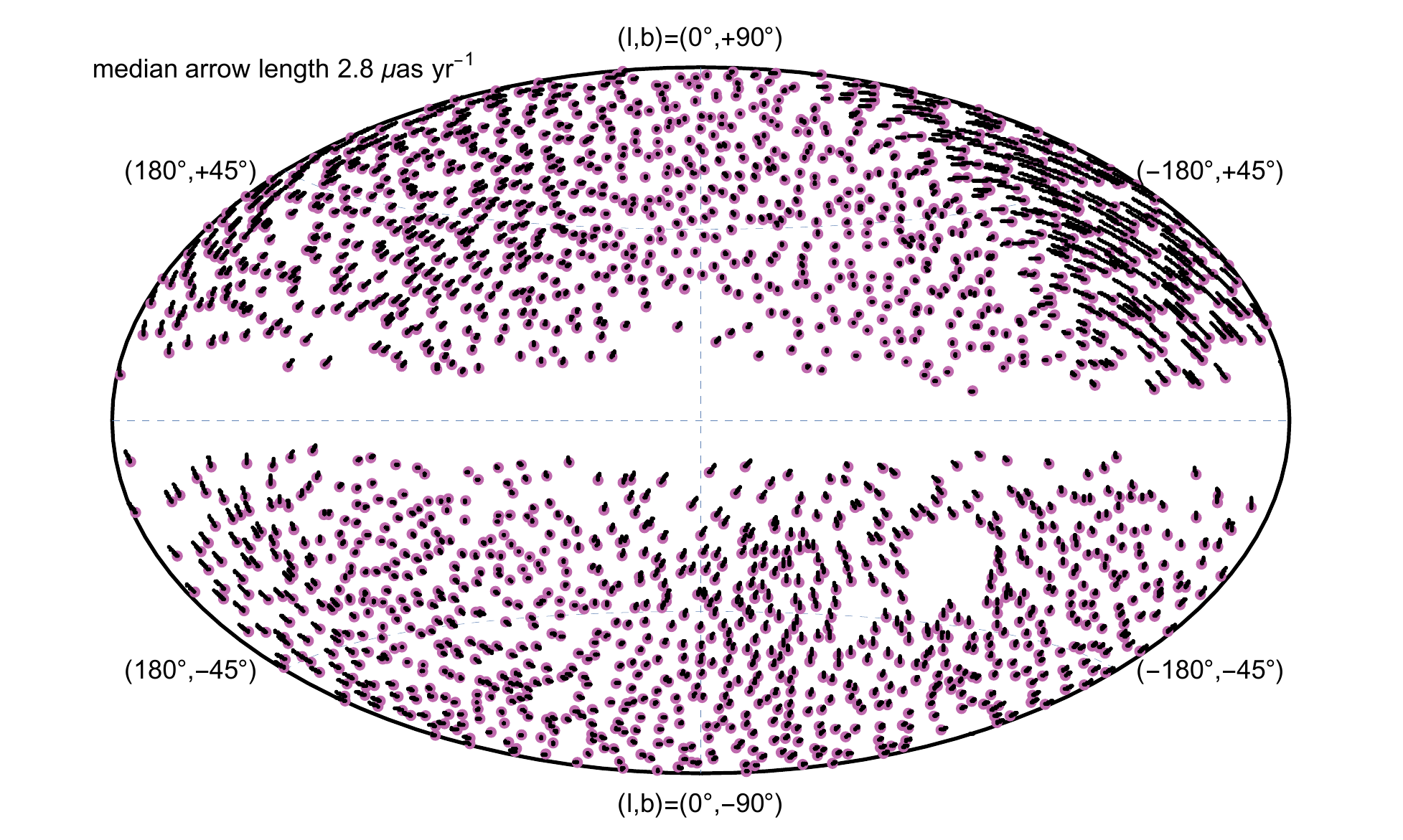}
   \caption{The differential ``corrected$-$original" VSH-fitted proper motion field of the same sample of Gaia CRF quasars as shown in Fig.\ref{field1.fig}. Purple dots at the origin of vectors indicate the mean positions of sources. The vectors are the changes in the fit induced by the conditional zero-parallax correction. }
    \label{field2.fig}
\end{figure}

Instead of the general pattern of the second VSH field, which is rather similar to Fig. \ref{field1.fig}, the direct difference between the two fits ``corrected$-$original" is displayed in Fig. \ref{field2.fig}. The updated scale of the arrows corresponds to the smaller magnitude of these differences. Comparing these two Figures, we conclude that the prominent south-north flow in the 2nd galactic quadrant becomes even more pronounced, also spreading to a large area in the northern part of the 3rd quadrant. This is truly the dominating feature, which is not consistent with the null model including a single secular aberration dipole. Although the streaming toward the Galactic center is more pronounced in the 2nd fit, the emergence of signal in \{ele,0,1,0\} also signifies that the 1st degree dipole is not as well aligned with this direction.

Although the updated proper motions bring the estimate of the aberration dipole closer to the expected value and the general Gaia estimate \citep{2021A&A...649A...9G}, this result should be considered with due caution. As is seen from Eq. \ref{mu.eq}, the corrections are defined by the standardized parallax measurement, whose probability density is symmetrically distributed around zero, and the signed correlations $r_{01}$ and $r_{02}$, while the impact of proper motion uncertainties is practically nullified by the optimal weights in the VSH adjustment. The sky-correlated pattern seen in Fig. \ref{field2.fig} is therefore defined by two seemingly independent patterns in the Gaia data: the sky distribution of the parallax offset from zero (parallax zero-point) and the ``average'' proper motion--parallax correlations as functions of sky coordinates. A dedicated experiment revealed that the tangential vector field $[r_{01},r_{02}](l,b)$ is indeed nonuniform showing large-scale deviations from zero in specific parts of the sphere. A VSH decomposition of this field with the same weights as the proper motion fitting produces a pattern closely resembling Fig. \ref{field2.fig} with a median magnitude of 0.179. Another numerical experiment with a tentatively corrected constant parallax offset revealed a modest degree of sensitivity of the VSH fit to this parameter. Thus, the sky distribution of proper motion updates mostly follows the peculiar distribution of astrometric correlations on the sky, and this application is validated only as long as the parallax offset has a significant accidental component. To quantify the \emph{sensitivity} of the low-degree terms to the adopted parallax zero-point, Table~\ref{zp.tab} compares the first-degree absolute amplitudes obtained with the original parallax as given in the catalog and with the object-specific SSH-based zero-point model of \citet{2026PASP..138i4503M}. Note that in this computation, the proper motions of 284,639  objects were updated only by the systematic offsets $\varpi_{\rm zp}^{\rm SSH}$ of individual parallax instead of the actual measured parallax in Eq.~\ref{mu.eq}. The low-degree coefficients shift by 1--2 of their formal errors (with the exception of the polar dipole), indicating that the VSH fit is moderately sensitive to the choice of zero-point model. This zero-point-only correction is distinct from the full conditional zero-parallax correction described earlier in this section: the latter uses the measured parallax of each source and raises the galactocentric dipole to $5.59$~\uasyr, whereas subtracting the deterministic SSH offset alone yields $4.91$~\uasyr. The two computations address different questions, which accounts for the difference in the recovered amplitudes.

\begin{table}[ht]
\centering
\caption{Sensitivity of first-degree VSH amplitudes in \uasyr\ (not to be confused with nominal VSH coefficients) to the adopted parallax zero-point. The formal errors in the same units from the VSH adjustment are shown in the fourth column; for reference, the original fit gives a dipole amplitude of $3.35$~\uasyr toward the Galactic center, while the zero-point-corrected fit produces $4.91$~\uasyr with a formal error of 0.72~\uasyr for both estimates. The normalization coefficients and nomenclature are listed in \citet{2025NatAs...9.1396M}. The second numerical column applies only the deterministic SSH zero-point contribution $\varpi_{\rm zp}^{\rm SSH}$, not the full measured-parallax conditioning of Eq.~\ref{mu.eq}.}
\label{zp.tab}
\begin{tabular}{lccc}
\hline\hline
Amplitude  & Original $\varpi$ & $\varpi-\varpi_{\rm zp}^{\rm SSH}$ & formal error \\
\hline
\{ele,0,1,0\} polar dipole (most changed) & $0.48$ & $2.96$ & $0.87$ \\
\{ele,1,1,1\} galactocentric dipole & $3.35$ & $4.91$ & $0.72$ \\
\{ele,2,1,1\} transverse dipole & $0.41$ & $0.83$ & $0.79$ \\
\{mag,0,1,0\} polar spin & $2.66$ & $1.93$ & $0.84$ \\
\{mag,1,1,1\} galactocentric spin & $0.86$ & $0.76$ & $0.65$ \\
\{mag,2,1,1\} transverse spin & $4.11$ & $4.80$ & $0.84$ \\
\hline
\end{tabular}
\end{table}

\section{Discussion}
Precision of Gaia CRF proper motions can be probabilistically improved using the strong prior of zero parallax for these extremely distant sources. The improvement can be quantified as a 15\% reduction of statistical outliers, for example. Although the majority of zero-parallax corrections is numerically quite small, some of the updates are comparable in absolute value to the given formal errors of proper motions. No external astrometric information is required, although the result remains conditional on the adopted zero-parallax prior and the assumed per-source Gaussian covariance model. On the mathematical side, the method has solid foundations in the classical frequentist theory of statistical inference.

The prime purpose of Gaia proper motions is to determine and subtract the rigid spin of the entire observed sample, which is arbitrary in an unconstrained global solution. It may be suggested that the a posteriori corrected values would be a better choice for this procedure. 
The inferred frame spin may be sensitive to the observed distribution of quasar parallaxes and to the sky-
dependent parallax–proper-motion correlation pattern.

As discussed in Sect. \ref{3.sect}, the impact of the conditional improvement of CRF proper motions in the low-degree VSH terms, i.e., the patterns of the largest angular scale on the sphere, depends on the proportion of the stochastic component in the low-degree scalar spherical harmonic (SSH) decomposition of the residual parallax field. In a separate analysis (Makarov \& Berghea 2026)\footnote{The SSH-based parallax-correction code (\texttt{varpi3.py}) and fitted coefficients are publicly available: V.~V.~Makarov \& C.~T.~Berghea 2026, \textit{Gaia parallax bias via spherical harmonics: \texttt{varpi3.py} --- a Python tool using Gaia DR3 data}, Zenodo, \url{https://doi.org/10.5281/zenodo.21708614}.}, an 8-degree SSH fit (81 fitting functions) was computed on 1.157 million Gaia DR3 parallaxes of carefully filtered CRF sources and analyzed as a function of color. Compared to the previously explored schemes \citep{2021A&A...649A...4L, 2024A&A...691A..81D}, a global SSH fit maximizes the benefit of the large statistical sample and allows us to extrapolate the model to the Galactic plane area. Only the constant SSH term $Y_{00}$ was found to be magnitude-dependent within a range of $-20$ to $0$ \uas. The weak dependence of the other 80 position-dependent terms on magnitude suggests that their origin may be stochastic, while the constant bias or offset is deterministic and instrumental as predicted by the preliminary error analysis \citep{2017A&A...603A..45B}. The instrumental bias is then propagated via the conditional updates in Eq. \ref{mu.eq}
to the corrected proper motion field. The adverse effect of this "overcorrection" is reduced, however, by the fact that the correlations $r_{01}$ and $r_{02}$ are commonly small and symmetrically distributed around zero. Indeed, half of the proper motion corrections from a general shift of all CRF parallaxes by 20 \uas\ are less than 3 \uasyr\ in absolute value, and 90\% are less than 9 \uasyr. The median formal error of CRF parallax is 513 \uas, while a quarter of the sample has parallax errors less than 322 \uas. Finally, a predefined model of parallax offset can be subtracted from the data prior to the conditional correction of proper motions for high-accuracy applications, e.g., replacing $\varpi$ with $\varpi-Z_5$ with the object-specific $Z_5$ from the model in \citep{2021A&A...649A...4L}.

The first-degree electric harmonics, which carry the secular aberration signal, are also sensitive to the conditional correction. Although the validity of the correction is not guaranteed in this case (conditioned by the possible accidental origin of the parallax bias), it may be prudent to compute both values to estimate the range of possible values. At the very least, the low-cost and easily implemented technique provides us a look at the hidden uncertainties of these fundamental determinations.

\section*{Appendix}

Let $\boldsymbol{v}$ be a multivariate normally distributed variable (vector), so that the probability density of $\boldsymbol{v}$ is ${\cal N}(\bar{\boldsymbol{v}}, \boldsymbol{\Sigma})$, where $\bar{\boldsymbol{v}}$ is the mean and $\boldsymbol{\Sigma}$ is the covariance matrix. Conditional probabilities emerge when the true mean values of some of the components of $\boldsymbol{v}$ are known a priori, so that we do not need to use the measurements or trials to infer them. The vectors and matrices can always be rearranged and partitioned in such a way that their leading portion includes the components with known means:
\eb 
\boldsymbol{v}=\begin{bmatrix}\boldsymbol{v}_1\\ \boldsymbol{v}_2\end{bmatrix}, \quad
\boldsymbol{\Sigma}=\begin{bmatrix}\boldsymbol{\Sigma}_{11} & \boldsymbol{\Sigma}_{12}\\
\boldsymbol{\Sigma}_{21} & \boldsymbol{\Sigma}_{22}\end{bmatrix},
\ee 
with the known mean $\bar{\boldsymbol{v}}_1$. 
The conditional probability of the unknown part $\boldsymbol{v}_2$ given a measured value $\boldsymbol{a}$ of $\boldsymbol{v}_1$ also follows a multivariate distribution with the conditional expectation \citep[e.g.,][]{seber}
\eb 
E(\boldsymbol{v}_2|\boldsymbol{a})=\bar{\boldsymbol{v}}_2+\boldsymbol{\Sigma}_{21} \boldsymbol{\Sigma}_{11}^{-1} (\boldsymbol{a}-\bar{\boldsymbol{v}}_1),
\label{e.eq}
\ee 
and the variance
\eb 
{\rm Var}(\boldsymbol{v}_2|\boldsymbol{a})=\boldsymbol{\Sigma}_{22}-\boldsymbol{\Sigma}_{21} \boldsymbol{\Sigma}_{11}^{-1} \boldsymbol{\Sigma}_{12}.
\label{var.eq}
\ee 
The proof can be found in \citep[][Exercise 2.9]{seber2003}. 
Using Eq. \ref{e.eq}, the unbiased sample estimate of the mean $\boldsymbol{v}_2$ is then
\eb 
\hat{\boldsymbol{v}}_2=\boldsymbol{b}-\boldsymbol{\Sigma}_{21} \boldsymbol{\Sigma}_{11}^{-1} (\boldsymbol{a}-\bar{\boldsymbol{v}}_1),
\ee 
where $\boldsymbol{b}$ is the sample (i.e., actual measurement) of $\boldsymbol{v}_2$. 

In the case of astrometric measurements considered in this paper, the vector $\boldsymbol{v}$ includes three variates: parallax and proper motion components. The prior information concerns the expectation of parallax, which can be safely assumed to equal zero. Thus, $\bar{v}_1=0$, $a$ is the measured value $\varpi$ given in the catalog, $\Sigma_{11}=\sigma_\varpi^2$, and Eq. \ref{e.eq} simplifies into Eq. \ref{mu.eq}.

Somewhat different derivations of Eqs. \ref{e.eq} and \ref{var.eq} can be found in a number of textbooks and internet resources, but we may benefit from a simple geometric interpretation of the conditional correction of measurement for the three-dimensional variate in hand. Let us arrange the three astrometric parameters of interest into a column vector $[x_0,x_1,x_2]^T$, so that $x_0$ corresponds to parallax, and $x_1$, $x_2$ are proper motion components. The formal covariance matrix
\eb 
\boldsymbol{\Sigma}= \begin{bmatrix}
\sigma_{0}^2 & r_{01}\sigma_{0}\sigma_{1} & r_{02}\sigma_{0}\sigma_{2}\\
r_{01}\sigma_{0}\sigma_{1} & \sigma_{1}^2 & r_{12}\sigma_{1}\sigma_{2}\\
r_{02}\sigma_{0}\sigma_{2} & r_{12}\sigma_{1}\sigma_{2} & \sigma_{2}^2 
\end{bmatrix}
\ee 
is given in the catalog, and it defines a 3D ellipsoid, which, in analogy with the 2D error ellipse \citep{1996ore}, can be called an error ellipsoid. The equation $\boldsymbol{x}^T\,\boldsymbol{\Sigma}^{-1}\,\boldsymbol{x}=C^2$ is the surface of equal probability density, and the volume enclosed by it is a specific confidence area. Since the correlation coefficients $r_{ij}$ are generally nonzero, the principal axes of the error ellipsoid are not aligned with the coordinate axes. The tilt results in the following effect: if the true value of parallax is zero, but the measurement is $\varpi$, the cross-section of the ellipsoid with a plane $x_0=\varpi$, which defines the actual error ellipse for $[x_1,x_2]^T$, is shifted from the true mean value of proper motion vector by a certain amount $[\Delta_1,\Delta_2]^T$. This conditional shift vector can be computed from the cross-section equation (where $x_0$ is replaced with $\varpi$) by perturbing the proper motion components $x_1\rightarrow x_1+\Delta_1$, $x_2\rightarrow x_2+\Delta_2$ and nullifying the terms linear with $\varpi$, $\Delta_1$, and $\Delta_2$. The approach is equivalent to finding an affine transformation (i.e., a translation of the coordinate system) that makes the cross-section a centered 2D ellipse. The resulting system of equations is
\begin{eqnarray}
    \varpi\,\sigma_1\sigma_2(r_{01}-r_{02}r_{12})&=& \sigma_0\sigma_2(1-r_{02}^2)\Delta_1+\sigma_0\sigma_1(r_{01}r_{02}-r_{12})\Delta_2 \nonumber\\
    \varpi\,\sigma_1\sigma_2(r_{02}-r_{01}r_{12})&=& \sigma_0\sigma_1(1-r_{01}^2)\Delta_2+\sigma_0\sigma_2(r_{01}r_{02}-r_{12})\Delta_1
\end{eqnarray}
which has a solution
\begin{eqnarray}
    \Delta_1&=&\frac{\varpi}{\sigma_0}r_{01}\sigma_1 \nonumber\\
    \Delta_2&=&\frac{\varpi}{\sigma_0}r_{02}\sigma_2.
\end{eqnarray}
These values should be subtracted from the measured proper motion components to equalize the expectancy with the true mean vector, which brings up Eq. \ref{mu.eq}.

PAS

\bibliography{main}{}
\bibliographystyle{aasjournalv7}

\end{document}